\documentstyle[12pt,aaspp4]{article}

\begin{document}

\title{$BVI$ CCD Photometry of the Globular Cluster 
47~Tuc\footnote{Based on observations collected at Las 
Campanas Observatory of the Carnegie Institution of Washington}}

\author{J. Kaluzny \& A. Wysocka}
\affil{Warsaw University Observatory, Al. Ujazdowskie 4,
00--478 Warszawa, Poland} 
\affil{\tt e-mail: jka@sirius.astrouw.edu.pl, wysockaa@sirius.astrouw.edu.pl} 

\author{K. Z. Stanek\altaffilmark{2}} 
\affil{Harvard-Smithsonian Center for Astrophysics, 60 Garden St., MS20, 
Cambridge, MA~02138} 
\affil{\tt e-mail: kstanek@cfa.harvard.edu} 
\altaffiltext{2}{On leave from N.~Copernicus Astronomical Center, 
Bartycka 18, Warszawa 00--716, Poland} 

\author{W. Krzemi\'nski}
\affil{Carnegie Observatories, Las 
Campanas Observatory, La Serena, Casilla 601, Chile}
\affil{\tt e-mail: wojtek@roses.ctio.noao.edu}

\begin{abstract}

We present $BVI$ photometry for about 16000 stars from a 220~arcmin$^{2}$
field centered 8~arcmin east of the center of 47~Tuc. We have identified
eight likely blue stragglers located in the outer parts of the cluster.
Four of these objects are easy targets for spectroscopic studies with
ground-based telescopes.  Six candidates for blue horizontal branch stars
were identified.  However, it is possible that all or most of them belong
in fact to the SMC halo. One faint blue star being candidate for a
cataclysmic variable was found close to the cluster center.

The average $I$-band magnitude for stars forming the red giant branch clump
is determined at $I_{0}=13.09\pm 0.005\;$mag.  This in turn implies
distance modulus of the cluster $(m-M)_{0,47Tuc}=13.32\pm 0.03\pm
0.036\;$mag (statistical plus systematic error), if we adopt $M_{I,m}=-0.23
\pm 0.03\;$mag for the average absolute luminosity of {\em
Hipparcos}-calibrated clump giants, following Paczy\'nski \& Stanek and
Stanek \& Garnavich. This distance modulus of 47~Tuc is lower by
$0.2-0.25\;$mag than its recent estimates based on {\em Hipparcos}\/
parallaxes for subdwarfs.  We discuss possible reasons for this
discrepancy. The photometric data are available through the {\tt anonymous
ftp} service.

\end{abstract}

\section{ Introduction}

47~Tuc (NGC~104) is a populous, metal rich, globular cluster located at
relatively small distance from the Sun. In the recent compilation Harris
(1996) lists for it metallicity ${\rm [Fe/H]}=-0.76$, absolute visual
magnitude $M_{\rm V}=-9.26$, distance modulus $(m-M)_{\rm V}=13.21$ and
extinction $E(B-V)=0.04$.  The cluster belongs to the population of disk
clusters (Zinn and West 1984).  High galactic latitude of 47~Tuc
($b=-44.9$) along with its closeness make it particularly attractive object
for detailed studies. The major, CCD based, photometric investigations of
47~Tuc include $BV$ photometry by Hesser et al. (1987) and $BVI$ photometry
by Alcaino \& Liller (1987). The cluster was a subject of numerous projects
conducted with the help of {\em HST}. The most recent contributions include
identification of a population of eclipsing binary stars in the cluster
center (Edmonds et al.~1996; Minniti et al.~1997), search for cataclysmic
binaries (Shara et al.~1996), study of the luminosity function (Santiago et
al.~1996) and spectroscopic study of a blue straggler located in the
cluster center (Shara, Saffer \& Livio 1997).  Kaluzny et al.~(1997, 1998)
used ground-based data to identify several new variables, including 12
eclipsing binaries, in the outer parts of the cluster.

In this paper we report medium-deep $BVI$ photometry for an eastern section
of 47~Tuc. These observations were aimed at: a) derivation of a luminosity
function for an upper main-sequence stars and subgiants; b) identification
of blue stragglers in the outer parts of the cluster; c) search for
candidates for blue horizontal branch stars.  The luminosity function
derived from our data will be published elsewhere (Wysocka 1998, in
preparation). Below we concentrate on search for blue stragglers and other
blue stars as well as on derivation of the distance modulus of the cluster.

\section{Observations and photometric reductions}

\begin{figure}[t]
\plotfiddle{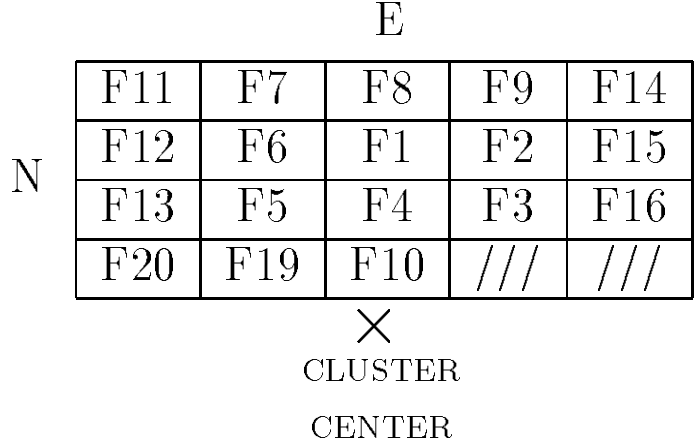}{5cm}{0}{150}{150}{-475}{-870}
\caption{A schematic finding chart showing location of
observed sub-fields relatively to the cluster center.}
\label{fig:find}
\end{figure}

The observations presented here were obtained using 1024$^{2}$ Tektronix
CCD (TEK2 camera) attached to the 2.5m duPont telescope of the Las Campanas
Observatory. The field of view of the camera was $4\times4$~arcmin$^{2}$
with the scale of 0.235 arcsec/pixel. The data were collected on 4 nights
of July 11--14, 1993. 18 partly overlapping fields covering eastern part of
the cluster were imaged through $BVI$ filters. A schematic chart showing
location of these fields is shown in Fig.~1. The total surface of the
observed area of the cluster equals about 220~arcmin$^{2}$. The following
sequence of exposures was collected for each sub-field: $B$-600~s,
$V$-300~s, $I$-300~s.  Additionally, shorter exposures were obtained for
sub-fields F19 and F10.\footnote{The tables with the $BVI$ photometry and
equatorial coordinates for all stars detected are available in electronic
form from {\tt ftp://www.astro.princeton.edu/kaluzny/Globular/47Tuc\_BVI}
or {\tt ftp://www.astrouw.edu.pl/pub/jka/Globular/47Tuc\_BVI }} The
seeing ranged from 1.1 to 1.7 arcsec with a median value of about 1.5
arcsec.  The instrumental profile photometry was extracted using
Daophot/Allstar package (Stetson 1987). The point spread function varying
linearly with coordinates was applied. The instrumental $bvi$ photometry
derived for sub-fields F1-20 was transformed to the common instrumental
system defined by photometry of sub-field F4. The offsets of the zero
points of the photometry obtained for individual sub-fields were determined
on a base of stars from the overlapping regions of adjacent sub-fields.

\begin{figure}[t]
\plotfiddle{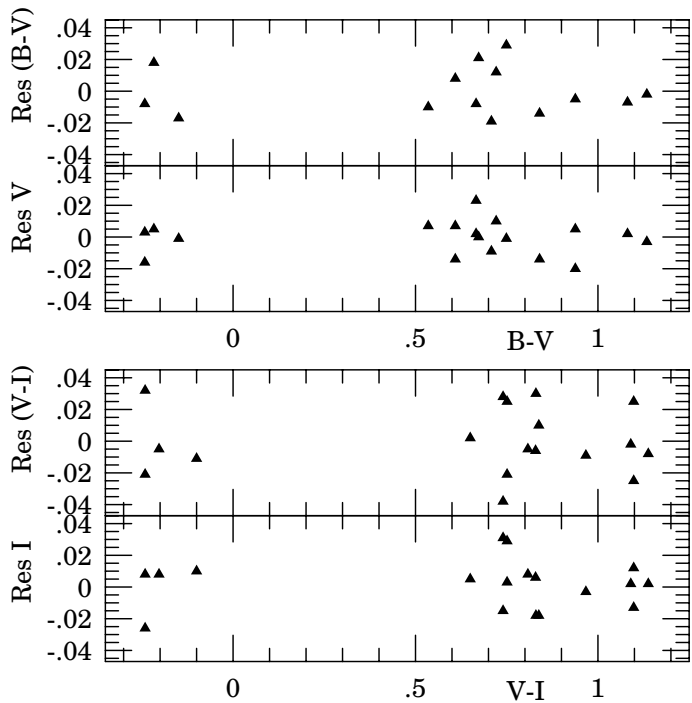}{8cm}{0}{110}{110}{-270}{-410}
\caption{Differences between our photometry of Landolt (1992)
standards and the published values.}
\label{fig:landolt}
\end{figure}

Calibration to the standard $BVI_{c}$ system was performed using photometry
of Landolt (1992) fields, observed over the whole run.  On the night of
July 12, 1993, we observed three Landolt fields containing a total of 14
standards. These fields were observed at air-masses spanning the range
$1.1-1.6$. Based on the aperture photometry of the standard stars we
adopted the following relations:

\begin{eqnarray}
v & = &1.698-0.024 \times(B-V) + 0.172\times(X-1.25) \nonumber \\ 
v & = &1.699-0.022 \times(V-I) + 0.170\times(X-1.25) \nonumber \\
i & = &1.953 - 0.017\times(V-I) +0.067\times (X-1.25) \\ 
b-v & = &0.394 +0.946 \times(B-V) +0.060\times(X-1.25) \nonumber \\ 
v-i & = &-0.247 +0.994 \times(V-I) + 0.083\times(X-1.25) \nonumber
\end{eqnarray}

where $X$ is an airmass and lower-case letters refer to the
instrumental magnitudes normalized to 1~s exposures. The overall
quality of the adopted  transformation is illustrated in Fig.~2 where
the magnitude and color residuals are plotted for the standard stars.
% Fig.~2 a) /STD/v.res  b) bv.res c) vis.res
The last step was determination of aperture corrections for the
sub-field F4. We estimate that probable errors of the zero points
of our photometry of 47~Tuc do not exceed $0.03\;$mag for magnitudes and
colors.  

\begin{figure}[p]
\plotfiddle{fig3.ps}{7.5cm}{0}{100}{100}{-320}{-300}
\caption{$V/B-V$ (left), $V/V-I$ (center) and $I/V-I$
color-magnitude diagrams for all observed sub-fields but F10 and
F19. Note the different vertical scale for the right panel.}
\label{fig:cmd1}
\plotfiddle{fig4.ps}{8.5cm}{0}{100}{100}{-320}{-320}
\caption{$V/B-V$ (left), $V/V-I$ (center) and $I/V-I$ color-magnitude 
diagrams for sub-field F19.}
\label{fig:cmd2}
\end{figure}
\begin{figure}[t]
\plotfiddle{fig5.ps}{7.5cm}{0}{100}{100}{-320}{-315}
\caption{$V/B-V$ (left), $V/V-I$ (center) and $I/V-I$ color-magnitude 
diagrams for sub-field F10.}
\label{fig:cmd3}
\end{figure}

The derived $BVI$ photometry of 47~Tuc is presented in the form of
color-magnitude diagrams (CMDs) in Figs.~3, 4 and 5.  Fig.~3 shows data for
all but two most crowded sub-fields F19 and F10.  Photometry for sub-fields
F19 and F10 is shown in Figs.~4 and 5, respectively.  Presented CMD's are
not complete in the sense that for each of sub-fields objects with
unusually large errors of photometry for their magnitude were discarded.

\subsection{Comparison with previous photometry}

The cluster region observed by us neither overlaps with the fields observed
by Hesser et el. (1987) nor with the field observed by Alcaino \& Liller
(1987). Hence it is impossible to compare directly our photometry with the
photometry published by these authors, but indirect comparison is possible.
In Fig.~6 we show $V/B-V$ CMD based on our data for the sub-field F4 with
over-imposed fiducial relation published by Hesser et al.~(1987; Table IX
in their paper). It has to be noted that parts of the fiducial relation
describing location of the horizontal branch and main sequence are based on
CCD data while the subgiants and lower giant branch are defined based on
the photographic data.  The overall agreement is good although a systematic
discrepancy in color is visible at the bottom of the subgiant branch.  The
average color for RGB stars plotted in Fig.~3 is $<B-V>=0.796$ which is
marginally bluer than the average color measured by Hesser et al.~(1987)
who obtained $<B-V>=0.800$ and $<B-V>=0.805$ for their fields F3 and F4.

\begin{figure}[p]
\plotfiddle{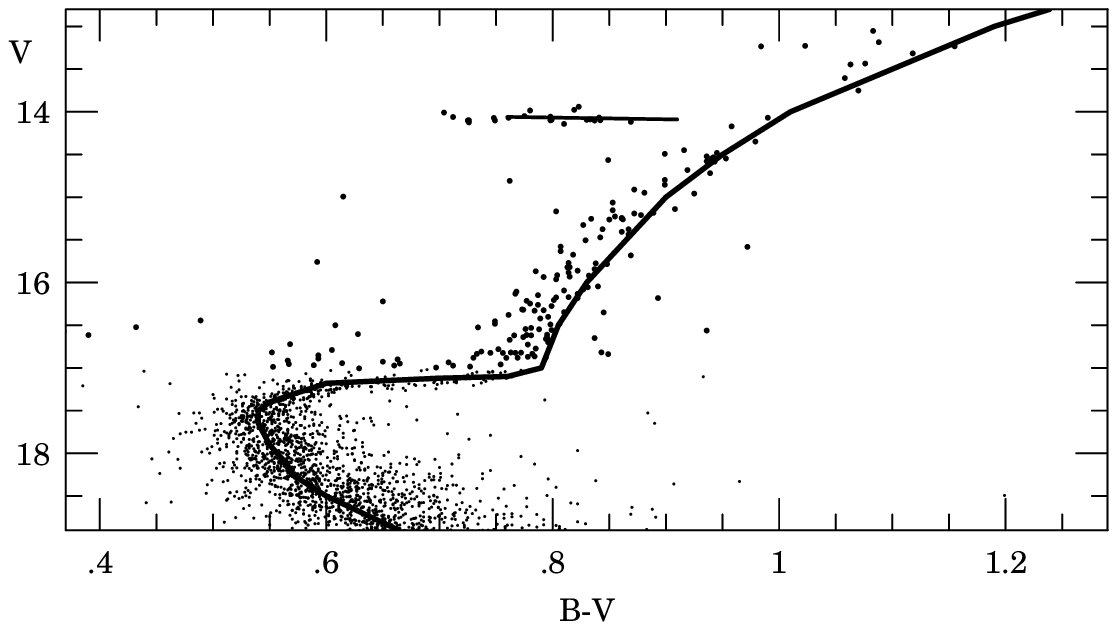}{7cm}{0}{120}{120}{-350}{-375}
\caption{The CMD for the sub-field F4, compared with the fiducial relation
obtained by Hesser et al.~(1987).}
\label{fig:heser}
\plotfiddle{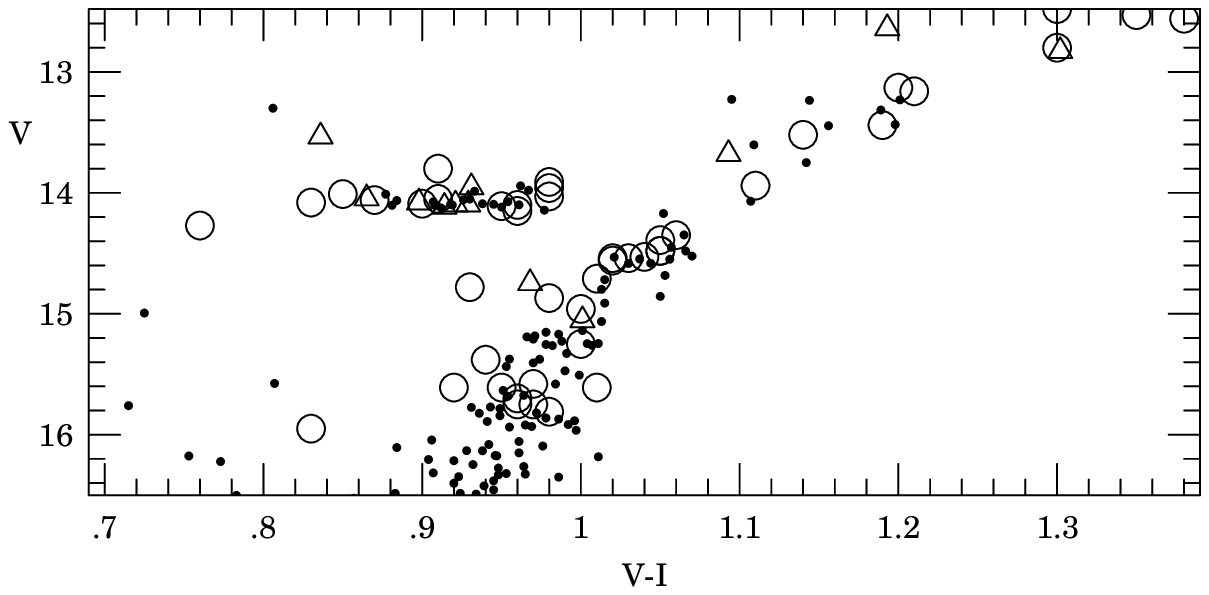}{7cm}{0}{120}{120}{-360}{-375}
\caption{Comparison of our $VI$ data (dots) with Armandroff
(1988; open circles) and Da Costa \& Armandroff (1990; open triangles).}
\label{fig:comp}
\end{figure}

Our photometry for main-sequence stars is consistent with results of
Alcaino \& Liller (1987) who obtained $(B-V)_{t}=0.56\pm 0.02$ and
$(V-I)_{t}=0.68\pm 0.02$ for the colors of the main-sequence turnoff. Our
data imply $(B-V)_{t}\approx 0.55$ and $(V-I)_{t}\approx 0.71$.

In Fig.~7 we compare our $VI$ photometry for RGB and sub-giant branch with
the data from Armandroff (1988) and Da Costa \& Armandroff (1990).  There
are no noticeable systematic differences between these 3 sets of data.

\section{Analysis of photometry}

\subsection{Blue stragglers}

Most of the thoroughly investigated globular clusters are known to harbor
more or less rich populations of blue stragglers (BSS) (eg. Ferraro, Fusi
Pecci \& Bellazzini 1995) There seems to be a correlation between a given
cluster concentration and the properties of its BSS population.  In sparse
clusters a substantial fraction of their BSS can be found in peripheral
regions, although generally BSS are more concentrated than "ordinary"
member stars (eg. subdwarfs). In clusters with very high central densities
BSS have been observed almost exclusively in their central
regions. Particularly interesting is the case of the nearby cluster M3
where there is an evidence for presence of two populations of BSS of
different origin (Ferraro et al.~1997). It has been suggested that BSS
observed in cores of globular clusters are formed preferentially by stellar
collisions while BSS observed in loose clusters are formed by merging of
primordial binaries (Ferraro, Fusi Pecci \& Bellazzini 1995; Sandquist,
Bolte \& Hernquist 1997).

Paresce et al.~(1991) reported identification of 22 BSS in the core of
47~Tuc, the first detection of BSS in a globular cluster core by {\em
HST}. In contrast Hesser al.~(1987) noted the complete lack of possible BSS
in the out-of-core regions the cluster.  Recently Kaluzny et al.~(1998)
surveyed a large fraction of the cluster and detected 12 eclipsing binaries
of which most are likely BSS.  Several candidates for BSS are visible in
the CMD's presented in Figs. 3, 4 and 5.  As photometry shown in Figs. 4
and 5 is relatively noisy we limited our attention to the data displayed in
Fig.~3.  An expanded view of that figure showing region around the
main-sequence turnoff is shown in Fig.~8.  One may note presence of an
apparent BSS clump at $V\approx 16.6$ and $B-V\approx 0.40$.  In Table 1 we
list photometry and equatorial coordinates for 8 stars forming that clump.
Star BSS7 has been identified as a contact binary by Kaluzny et al. (1998;
variable OGLEGC226).  Hence it is a likely binary BSS.  The finding charts
allowing identification of 8 identified BSS candidates are shown in Fig.~9.

\begin{figure}[p]
\plotfiddle{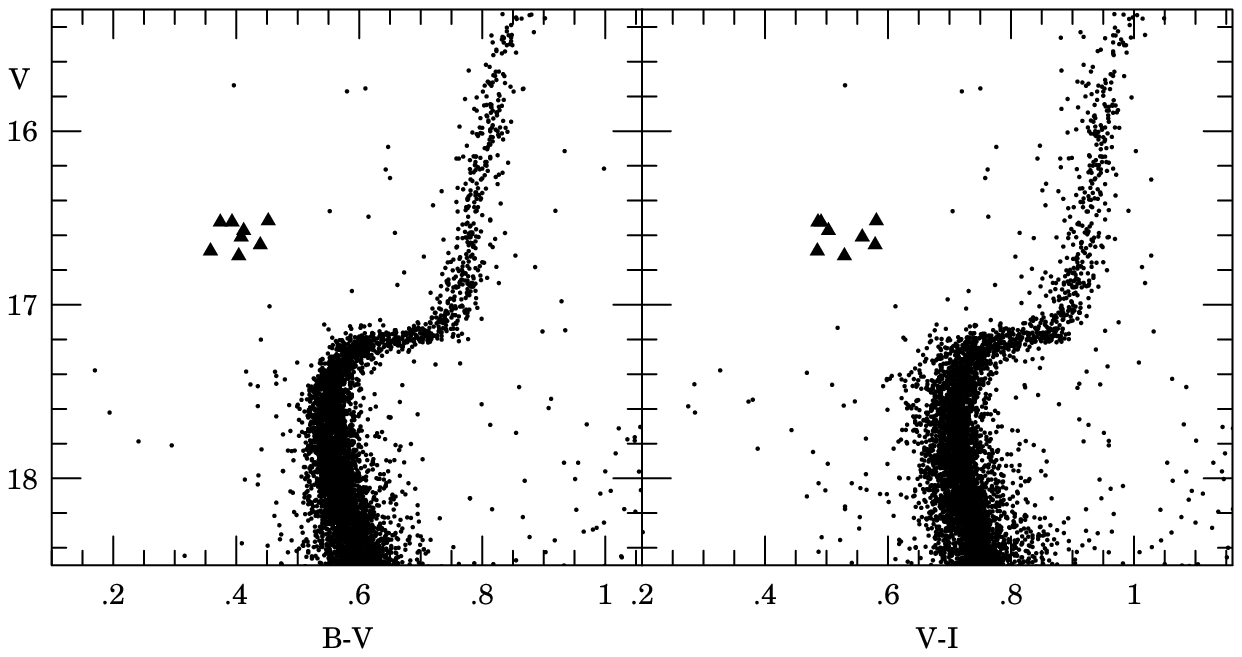}{6cm}{0}{120}{120}{-380}{-375}
\caption{An expanded view of the main-sequence turnoff region
for data presented in Fig.~3. Eight likely BSS are marked with
triangles.}
\label{fig:bss}
\plotfiddle{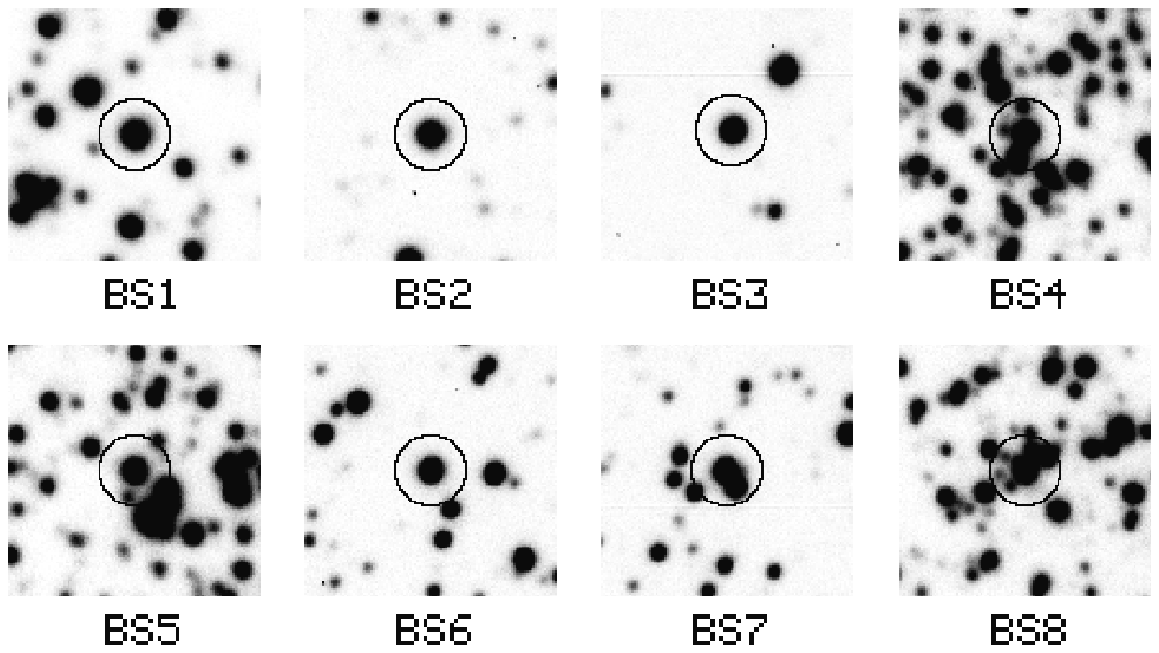}{8.5cm}{0}{120}{120}{-310}{-330}
\caption{The $V$-band finding charts for 8 BSS candidates. Each chart
is 60 arcsec on a side with north up and east to the left.}
\label{fig:findbss}
\end{figure}

Recently Shara, Saffer \& Livio (1997) used a spectrum obtained with the
{\em HST}\/ to derive absolute parameters for one of BSS located in the
cluster core.  We note in that context that four of our BSS candidates are
relatively isolated, un-crowded objects.  They are excellent objects for
spectroscopic studies with ground-based telescopes. As one of the closest
globular clusters 47~Tuc is particularly well suited for studies aimed at
comparison of properties of individual stars belonging to inner and outer
BSS populations.

\setcounter{table}{0}
\begin{table}
\begin{center}
\caption[]{Photometry and coordinates for blue straggler candidates}
\vspace{3mm}
\begin{tabular}{lllllc}
\hline
ID & {\it V} & {\it B-V} & {\it V-I} &  RA(1950) &  Dec(1950)\\
   &         &           &           & [h:m:sec] & [$^{\circ}$:':'']\\
\hline
BS1 & 16.528 & 0.449 & 0.578 & 0:23:26.9 & $-$72:14:20.4\\
BS2 & 16.534 & 0.389 & 0.481 & 0:24:43.9 & $-$72:19:38.5\\
BS3 & 16.535 & 0.369 & 0.486 & 0:24:57.6 & $-$72:14:10.8\\
BS4 & 16.584 & 0.409 & 0.498 & 0:23:03.1 & $-$72:22:42.5\\
BS5 & 16.621 & 0.404 & 0.555 & 0:23:07.5 & $-$72:22:24.6\\
BS6 & 16.666 & 0.436 & 0.576 & 0:23:51.4 & $-$72:26:52.6\\
BS7 & 16.701 & 0.352 & 0.480 & 0:24:00.4 & $-$72:27:42.6\\
BS8 & 16.729 & 0.400 & 0.525 & 0:23:03.4 & $-$72:17:24.1\\
\hline
\hline
\end{tabular}
\end{center}
\end{table}

\subsection{Distance modulus of the cluster}

Hesser et al.~(1987) derived the distance modulus of 47Tuc
$\mu_{0,47Tuc} = 13.25\;$mag. Recently Gratton et al.~(1997) derived a
higher value of $13.44\;$mag, using {\em Hipparcos} calibrated
subdwarfs and reddening of $E(B-V)=0.023\;$mag.  Reid (1998) also used
{\em Hipparcos} calibrated subdwarfs and obtained distance modulus of
$13.57\pm 0.15\;$mag, assuming $E(B-V)=0.04\;$mag (which would become
$13.50\;$mag had he used Gratton et al.'s value of reddening).
Finally, Salaris \& Weiss (1998) obtained a distance modulus of
$13.50\pm 0.05\;$mag, assuming reddening of $E(B-V)=0.05\;$mag.

Following an approach developed by Paczy\'nski \& Stanek (1998), here
we derive a distance to 47Tuc by comparing red clump stars from the
{\em Hipparcos}\/ catalog with the red clump stars from our data
set. This method was applied by Stanek \& Garnavich (1998) to M31
galaxy, by Udalski et al.~(1998) to the LMC and the SMC and by Stanek,
Zaritsky \& Harris (1998) to the LMC.

In the upper panel of Fig.~10 we show the red clump dominated part of
the $I_0,\; (V-I)_0$ color-magnitude diagram for 47Tuc, corrected for
the extinction and the reddening assuming a value of $E(B-V)=
0.04\;$mag used by Reid (1998).  The dashed rectangle corresponds to
the region of the CMD selected for comparison with the local red clump
stars observed by {\em Hipparcos} (Paczy\'nski \& Stanek 1998).  In
the lower panel of Fig. 10 we show the distribution of 138 stars from
the selected region as a function of extinction-corrected magnitude
$I_0$.  Following Stanek \& Garnavich (1998), we fitted this
distribution with a function
\begin{equation}
n(I_0) = a + b (I_0-I_{0,m})  + c (I_0-I_{0,m})^2 +
\frac{N_{RC}}{\sigma_{RC}\sqrt{2\pi}}
 \exp\left[-\frac{(I_0-I_{0,m})^2}{2\sigma_{RC}^2} \;\right].
\end{equation}
The first three terms describe a fit to the ``background''
distribution of the red giant stars, and the Gaussian term represents
a fit to the red clump itself.  $I_{0,m}$ corresponds to the peak
magnitude of the red clump population. We obtained the value of
$I_{0,m}=13.089 \pm0.005\;$mag and $\sigma_{RC}=0.026\;$mag, i.e.
extremely narrow, well defined red clump.

\begin{figure}[t]
\plotfiddle{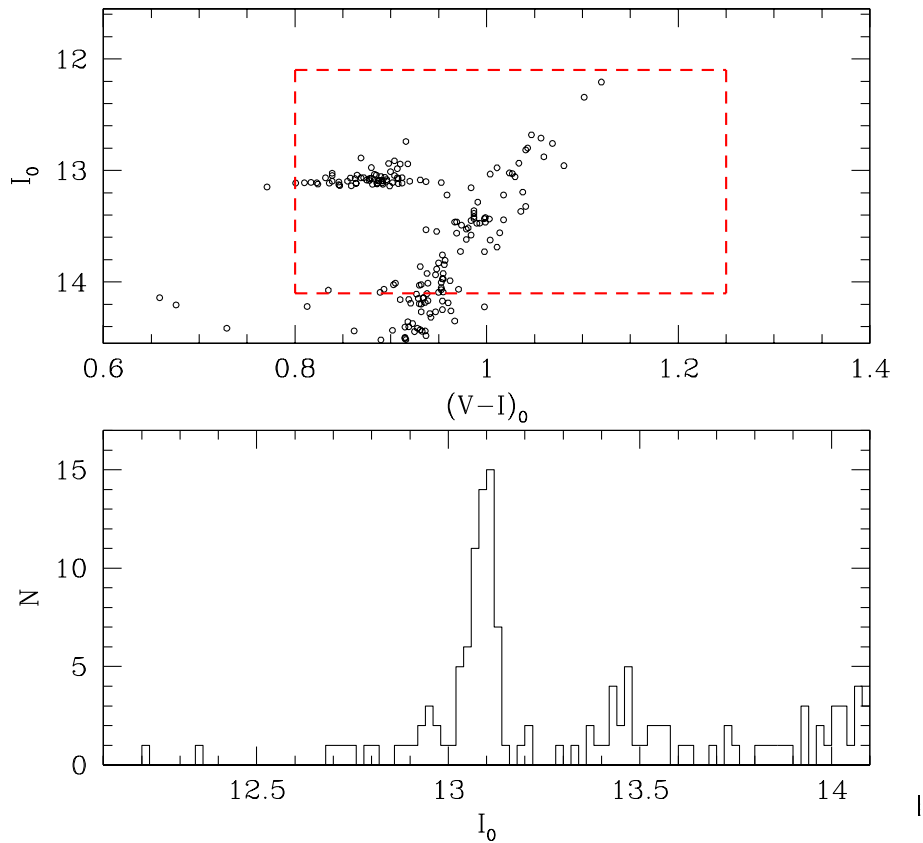}{9.5cm}{0}{60}{60}{-200}{-110}
\caption{The red clump dominated parts of the $I_0,\; (V-I)_0$ CMD,
corrected for the extinction and the reddening assuming a value of
$E(B-V)= 0.04\;$mag (upper panel).  The dashed rectangles surround the
red clump region used for the comparison between the local (observed
by {\em Hipparcos}) and the LMC red clump stars. In the lower panel
we show the distribution of 138 stars from the selected region as a
function of extinction-corrected magnitude $I_0$.}
\label{fig:clump}
\end{figure}

We now proceed to obtain the 47Tuc distance modulus using the red
clump, by assuming that the absolute $I$-band brightness of the red
clump stars is the same for the local stars observed by {\em
Hipparcos}\/ and those in the cluster. Combining ${I}_{0,m}$ with the
distribution of local red clump stars, which have $M_{I,m}=-0.23
\pm0.03$ (Stanek \& Garnavich 1998), we obtain the distance modulus
for the 47Tuc, $\mu_{0,47Tuc} = 13.319 \pm 0.030\;$mag (statistical
error only). After adding the systematic error of $0.02\;$mag due to
the uncertainty in the extinction $A_I$ determination, and $0.03\;$mag
due to the zero-point uncertainty in our $I$-band photometry, we
arrive at the final value of $\mu_{0,47Tuc}= 13.319 \pm 0.030 \pm
0.036\;$mag (statistical plus systematic error). This is $\sim
0.2-0.25\;$mag below the value of Reid (1998) and Salaris \& Weiss
(1998), but in good agreement with the value of Hesser et al.~(1987).

One possible explanation for this difference is that the intrinsic
brightness of the red clump stars in 47Tuc is higher than in the Solar
neighborhood.  If following Reid (1998) we adopt $(m-M)_{0,47Tuc}=
13.57\;$mag for the cluster, then we obtain $M_{\rm I}=-0.48$ for clump
giants with $[{\rm Fe/H}]\approx -0.7$. However, note that the distance of
Reid (1998) has a high statistical error of $0.15\;mag$, as it is based on
only nine stars. Our comparison is based on about $\sim 600$ {\em
Hipparcos} red clump stars and about $\sim 100$ red clump stars from
47Tuc. This shows the great potential of the red clump method as a distance
scale indicator.

\subsection{Blue stars}

It has been realized in recent years that some old stellar clusters
with apparently red horizontal branches possess in fact populations of
sdB/O stars.  The most striking example is an extremely
metal-rich old open cluster NGC~6791 in which about 30\% of helium
burning giants form an extended blue horizontal branch (Kaluzny \&
Udalski 1992, Kaluzny \& Rucinski 1995). More recently Rich et
al.~(1997) discovered extended blue horizontal branches in two
metal-rich globular clusters NGC~5927 and NGC~6388. Theoretical models
aimed at explaining formation of UV-bright helium-burning stars in old
metal-rich populations were discussed among others by Yi, Demarque \&
Kim (1997) and Sweigart \& Catelan (1998). Existence of hot
long-living stars in old metal-rich stellar systems may offer an
attractive explanation of excessive UV flux in some giant elliptical
galaxies (Code \& Welch 1979; Burnstein et al.~1988; Brown et
al.~1997).

The $V/B-V$ CMD shown in Fig.~3 includes 5 blue stars with $B-V\approx
-0.1$ and $18<V<18.7$. Photometry and coordinates of these objects are
listed in Table 2. That table includes also one relatively bright blue star
from Fig. 3 (star B6) and one faint blue star identified in the sub-field
F10 (star B7; see Fig. 5).  Finding charts for stars B1-7 are shown in
Fig. 11.  Admittedly stars B1-5 lie on the extension of the upper
main-sequence of the SMC which is clearly visible in the lower left part of
the $V/B-V$ CMD shown in Fig. 3. Luminosities and colors of stars B1-5 are
consistent with the hypothesis that they are dwarfs of spectral types B7-9
belonging to the SMC halo.  Note however, that there is a $\approx 1$~mag
gap between these 5 stars and fainter candidates for the upper-main
sequence stars from the SMC.  On the other hand, it is striking that CMD's
for sub-fields F10 and F19, which are located closest to the cluster
center, are void of candidates for sdB/O stars. Sub-fields F10 and F19
covered about $1/9$ of the total surveyed area. Their $V/B-V$ CMD's
presented in Figs. 4 and 5 include 2941 stars with $17<V<19$ while
corresponding number for the remaining sub-fields (see Fig. 3) is 6934.
Hence, it would be reasonable to expect in Figs. 4 and 5 presence of 1-3
blue stars with properties similar to stars B1-5

Star B6 is much brighter that stars B1-5. If it were the SMC member than
its absolute magnitude would be $M_{V}\approx -3.7$ what corresponds to a
dwarf of spectral type B0 or B1. However, the observed color of the star is
inconsistent with such hypothesis.

The CMDs presented in Figs.~3, 4 and 5 contain a total of 226 stars located
on the red horizontal branch of 47~Tuc. We have identified 6 candidates for
stars from extension of the blue horizontal branch, but 5 of them are
likely to be in fact upper main sequence stars from the SMC.  The question
about cluster membership of stars B1-6 can be easily resolved by measuring
radial velocities of these objects.  In any case, our results confirm
earlier findings about paucity or even lack of hot subdwarfs in 47~Tuc.

\begin{figure}[t]
\plotfiddle{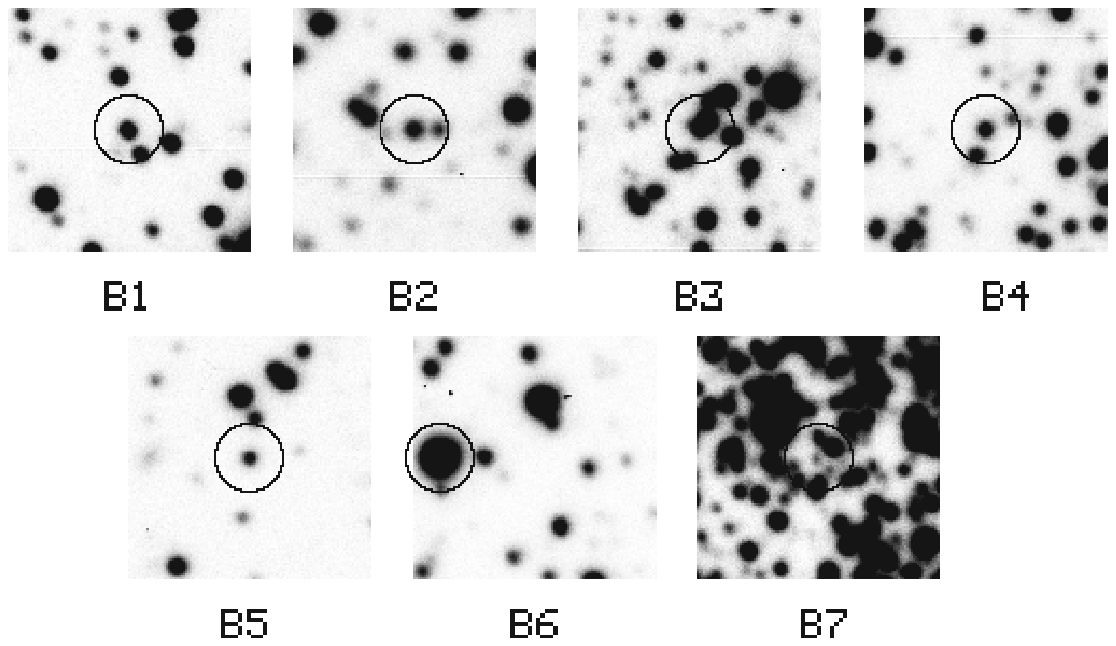}{6cm}{0}{120}{120}{-355}{-400}
\caption{The $V$-band finding charts for blue stars B1-B7 from Table 2.
Each chart is 60 arcsec on a side with north up and east to the left.}
\label{fig:findblue}
\end{figure}

\setcounter{table}{1}
\begin{table}
\begin{center}
\caption[]{Photometry and coordinates for blue stars identified in
the 47~Tuc field}
\vspace{3mm}
\begin{tabular}{llllcc}
\hline
ID & {\it V} & {\it B-V} & {\it V-I} &  RA(1950) &  Dec(1950)\\
   &         &           &           & [h:m:sec] & [$^{\circ}$:':'']\\
\hline
\hline
B1 &  17.998 & -0.053 & -0.024 & 0:23:49.6 & -72:24:15 \\
B2 &  18.159 & -0.135 & -0.162 & 0:24:03.5 & -72:20:34 \\
B3 &  18.172 & -0.154 & -0.079 & 0:23:18.5 & -72:23:32 \\
B4 &  18.577 & -0.132 & -0.156 & 0:22:32.2 & -72:13:56 \\
B5 &  18.657 & -0.120 & -0.225 & 0:24:49.9 & -72:19:47 \\
B6 &  15.040 & -0.064 & -0.042 & 0:23:36.7 & -72:14:45 \\
B7 &  20.226 & -0.224 &        & 0:22:40.7 & -72:20:04 \\
\hline
\hline
\end{tabular}
\end{center}
\end{table}

Object B7 is too faint to be a candidate for a hot subdwarf belonging to
47~Tuc. It was included in Table 2 because it is noticeably bluer than an
upper-main sequence stars from SMC which can be seen in Fig. 3.  Although
B7 is located in the densest surveyed sub-field its photometry is quite
reliable as luckily that star is a relatively un-crowded object (see
Fig. 11). The derived magnitude and color of B7 are consistent with a
hypothesis that it is a cataclysmic variable belonging to 47~Tuc.  So far
only one cataclysmic variable was identified in the cluster (Paresce \& De
Marchi 1994; see also Shara et al. 1996).  Given an unexpected paucity of
cataclysmic variables in globular clusters (eg. Shara et al. 1996) it may
be worth to examine closer nature of B7. The star can be monitored for
possible variability and also its spectrum can be acquired with the
ground-based telescopes.

\section{Conclusions}

We summarize our results as follows:

\begin{enumerate}

\item{} We have identified 8 promising candidates for blue stragglers
in the outer parts of 47~Tuc. Four of these stars are uncrowded objects
being easy targets for detailed spectroscopic studies with ground-based
telescopes.

\item{} Five candidates for stars from the extended horizontal branch
of 47~Tuc were identified in the surveyed field. However, these objects may
be as well B-type main sequence stars belonging to the halo of the SMC. One
relatively bright object being a candidate for blue horizontal branch star
was also found.

\item{} A faint blue star was identified close to the cluster center.
Its apparent magnitude and color are consistent with the hypothesis that it
is a cataclysmic variable belonging to the cluster.

\item{} The average $I$-band magnitude for stars forming the red giant
branch clump is determined at $I_{0}=13.09\pm 0.005\;$mag.  This in turn
implies distance modulus of the cluster $(m-M)_{0,47Tuc}=13.32\pm 0.03\pm
0.036\;$mag.  

\end{enumerate}

\acknowledgments{JK, AW and WK were supported by the Polish KBN grant
2P03D011.12.  Partial support was provided with the NSF grant AST-9528096
to B.~Paczy\'nski.  KZS was supported by the Harvard-Smithsonian Center for
Astrophysics Fellowship.}

\end{document}